# Two-phase xenon emission detector with electron multiplier and optical readout by multipixel avalanche Geiger photodiodes

**D.Yu. Akimov**[a,b], **A.V. Akindinov**[a], **I.S. Alexandrov**[a,b] [*], **V.A. Belov**[a,b], **A.I. Bolozdynya**[b], **A.A. Burenkov**[a,b], **A.F. Buzulutskov**[c], **M.V. Danilov**[a,b], **Yu. V. Efremenko**[b,d], **M.A. Kirsanov**[b], **A.G. Kovalenko**[a,b], **V.N. Stekhanov**[a,b]

[a] *State Scientific Centre of Russian Federation Institute for Theoretical and Experimental Physics (ITEP),*
*25 Bolshaya Cheremushkinskaya str., Moscow, Russian Federation*

[b] *National Nuclear Research University "MEPhI",*
*31 Kashirskoe highway, Moscow, Russian Federation*

[c] *Budker Institute of Nuclear Physics,*
*11 Lavrentiev ave., Novosibirsk, Russian Federation*

[d] *University of Tennessee,*
*1408 Circle Park Drive, Knoxville, Tennessee, USA*
*E-mail:* aleksandrov@itep.ru

ABSTRACT: The successful operation of a new optical readout system (THGEM + WLS + MGPDs (multichannel array of multipixel avalanche Geiger photodiodes) in a two-phase liquid xenon detector was demonstrated.

---

[*] Corresponding author

**Contents**

**1. Introduction**

In the last decade, liquid noble gas detectors became widely accepted in low background experiments to search for rare events. The example of successful application of such a technology is a two-phase emission detector for dark matter search. In such a detector, the sensitive part is a TPC (Time Projection Chamber) filled with a liquid noble gas, which serves as a target for detection of a dark matter candidates such as WIMPs (Weakly Interacting Massive Particles) (see reviews [1],[2],[3],[4],[5]). Excitations and ionisations are created in the massive liquid phase. The ionised electrons extracted into the gas phase generate an intense electroluminescence which is used to obtain X and Y coordinates (in a horizontal plane) and to measure the charge. The Z coordinate is reconstructed from the time difference between the scintillation and ionisation signals. This type of detector is also under consideration for another application which requires sensitivity in a sub-keV region of energy deposition, namely, the detection of coherent neutrino scattering off atomic nuclei [6],[7]. In this case, high sensitivity to a single ionisation electron is required. In the current generation of two-phase emission detectors, photomultipliers (PMTs) are used to readout the scintillation and electroluminescence light.

Several experimental groups, which develop noble gas detectors, have recently started to study the possibility of using new semiconductor devices – multipixel avalanche Geiger photodiodes (MGPD) as light collection detectors [8],[9],[10],[11],[12],[13],[14],[15]. The other commonly used brand names of this devise are SiPM, MPPC, MRS APD and GAPD. These photodetectors operate in a single photon counting mode and may replace photomultipliers in the future. Because of the very low mass they are expected to contribute an extremely low radioactive background in comparison to that of PMTs. This is very important in low-background experiments. Another advantage is their compact dimension. One can assemble a thin photosensitive array with a high granularity in order to obtain good coordinate resolution. The high space resolution is helpful in experiment on the detection of coherent neutrino scattering off atomic nuclei to suppress a spontaneous noise of single electron emission in xenon two-phase detectors [16],[17]. The dramatic decrease of the photodiode noise at cryogenic temperatures (by several orders of magnitude) in comparison to the noise level at the room temperature favours the use of MGPDs for readout in liquid noble gas detectors. The use of a thick gas electron multiplier (THGEM [18]) in the gas phase of a two-phase detector in front of an array of MGPDs can provide additional light amplification and therefore significantly improve the sensitivity of a two-phase emission detector a for small energy depositions. Such structures are called Cryogenic Avalanche Detectors (CRADs): see review [19] and its references.

Unfortunately, till now there are no commercially available MGPDs sensitive to Vacuum Ultra Violet (VUV) light in the emission region of noble gases (175 nm for Xe). In our previous study [12], we have shown that a wavelength shifter (WLS) specially developed for operation in the liquid xenon (LXe) environment (p-terphenyl on a sapphire plate coated with a poly-para-xylilen protection film) can provide the overall photon detection efficiency (PDE) of MGPD to the 175-nm xenon emission at a level of ~ 10%.

It should be noted that the optical readout of the THGEM multiplier in two-phase CRADs in Ar and gaseous CRAD in Xe has been successfully demonstrated either in the VUV [20] or Near Infrared [21],[22] spectral range, using a single optical channel, namely a single MGPD.

The aim of this work is first the experimental demonstration of the operation of a multichannel optical readout system based on THGEM, a wavelength shifter, and an array of 19 MGPDs in a two-phase xenon detector.

## 2. Experimental setup.

### 2.1. Detector.

We performed our tests in a two-phase liquid Xenon detector, which was previously used as a full-functional small prototype of the ZEPLIN-III dark matter detector [23],[24]. The design of the electrode structure is similar to that in the ZEPLIN-III, but the diameter is smaller (105 mm).

The main elements of the construction are: an array of MGPDs (1 in figure 1), a WLS coated sapphire window with poly-para-xylilen protection film (2), THGEM (3,4), an intermediate field-shaping ring (5), a wire cathode (6) and a wire protection screen of photocathodes (0.1-mm stainless steel wire; 1-mm pitch; 7), and an array of seven PMTs (8). The PMTs view the LXe volume from the bottom. The THGEM plate is installed instead of a mirror anode in the earlier version of the prototype. We've tested both single and double THGEM structures. In this paper, we present the results obtained with the double THGEM. The distance between the cathode and THGEM is 27 mm. The electrode structure is filled with the LXe to the level of 5 mm below the bottom THGEM metal plane. The distance between THGEMs was 3 mm. The distance from THGEM to the sapphire glass was 6 mm (9 mm for the case of single THGEM), and the distance from the WLS plane to the MGPDs was 3.2 mm.

The array of "blue sensitive" p-on-n MGPDs (by CPTA LTD, Russian Federation [25]; a vendor abbreviation is MRS APD) consists of 19 devices packed in a Teflon holder with a distance of 10 mm between them (see figure 2). The size of each MGPD is 2 x 2 mm, and each contains 1584 cells. The PDE of the MGPD at a maximum wavelength of the p-terphenyl emission (~370 nm) is more than 15% (curve 4 in figure 1 of [12]). The set of photodetectors was produced according to our requirements of cleanness for operation in the LXe medium: the front surfaces are left open (without any protection compound). Study of operation of MGPD of this type is presented in [12].

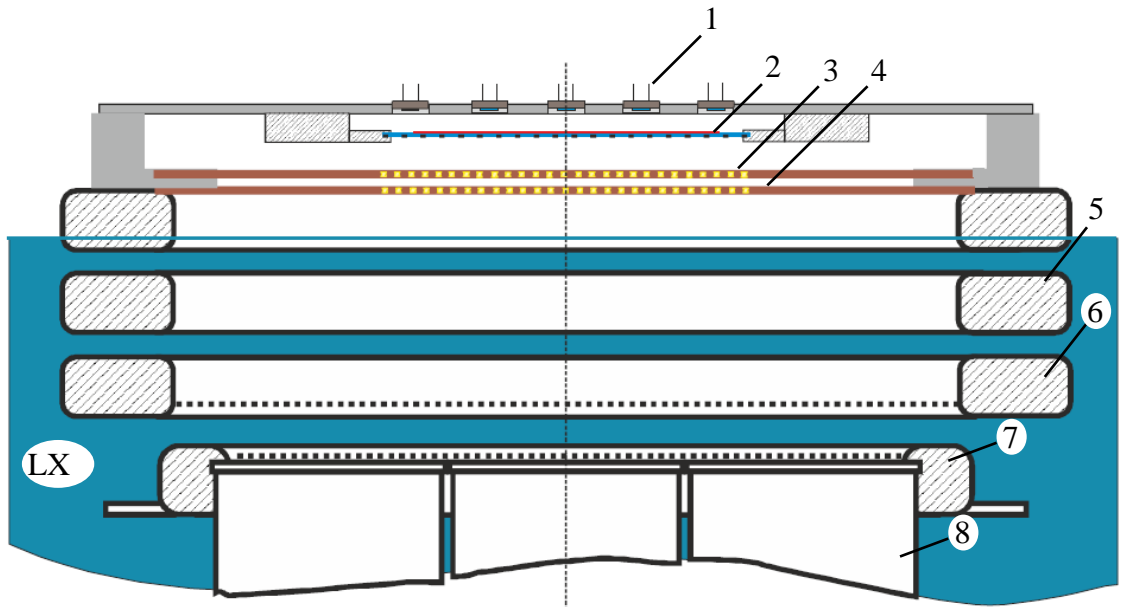

Figure 1. Schematic view of the two-phase detector. 1 – array of MGPDs, 2 – WLS on a sapphire window and with optically transparent metal mesh (on the bottom), 3 and 4 – THGEM2 and THGEM1, correspondently, 5 – intermediate field-shaping ring, 6 – cathode, 7 –photocathodes protection screen, 8 – array of seven PMTs.

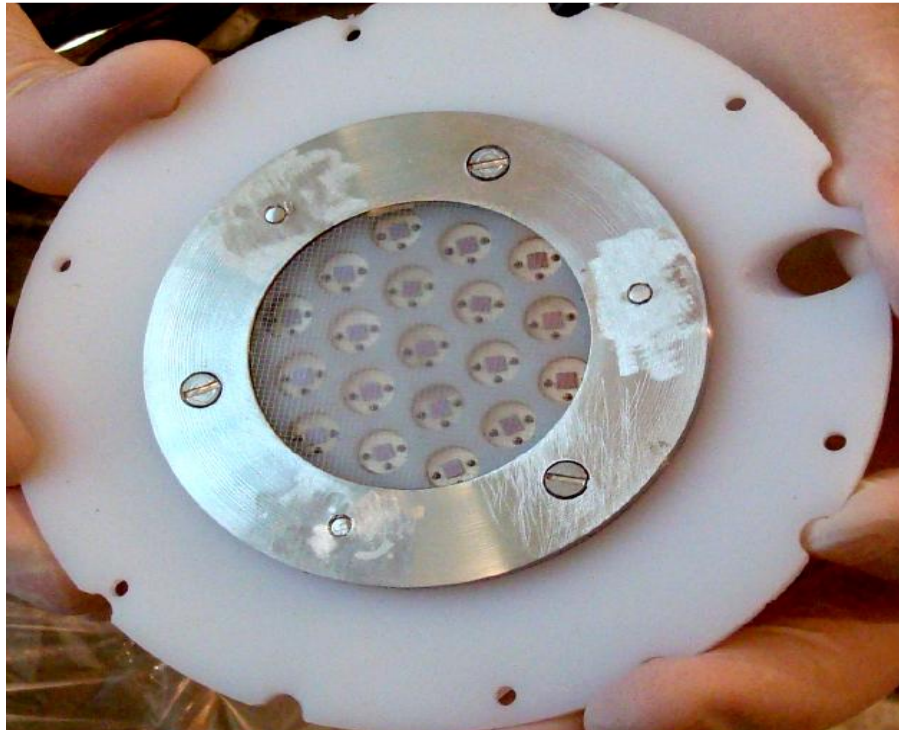

**Figure 2.** The photo of the MGPD photosensitive array assembled together with WLS on a sapphire window and with optically transparent metal mesh.

The wavelength shifting optical element (40-mm diam.) was produced by vacuum deposition of p-terphenyl on a sapphire VUV transparent disc with a diameter of 50 mm. The p-terphenyl layer was coated by a poly-para-xylilen film to protect the LXe medium from contamination by the volatile p-terphenyl molecules [26]. The disk was then oriented with the open surface (without WLS) to THGEM.

The PMTs are VUV-sensitive multialkali 30-mm FEU-181 with $MgF_2$ window, produced by MELS. The measured quantum efficiency of the photocathode at 200 nm is 15%. The PMT array has a common base placed outside the chamber.

The THGEM by CPTA LTD was made from 250-µm thick Kapton, coated by 35-µm copper layer on both sides. The holes have a diameter of 0.4 mm, the hole pitch is 0.7 mm with 0.1-mm rim at the edge of the holes on both sides. The active diameter of THGEM is 50 mm. We have tested both single and double THGEM structures. The HV potentials to the THGEMs were applied through the gaseous xenon by wires placed inside isolating Kapton pipes and were limited to values of several kV. In the original construction of the prototype there was not any grid under the LXe surface. Therefore, we had to apply the large negative potential to the cathode to ensure sufficient extraction efficiency of ionisation electrons from the liquid to the gas phase. Thus, we were significantly restricted in the total HV bias applied to the electrode system.

## 2.2. Signal readout and DAQ.

The signals from MGPDs were amplified with fast current amplifiers assembled in a breakout box mounted on a multi pin feed-through on the cryostat vacuum jacket. In this box, there was also a power supply distributor in order to provide individual operation voltage for each MGPD. The distributor was necessary because of the significant dispersion of a breakdown voltage for the given set of photodetectors. The operation voltages were set at 0.7 V above breakdown points. The signals from the 16 amplifiers (out of 19 of the full array) were digitized with two 8-channel 12-bit 200-MS/s VX1720 CAEN VME modules.

The signals from the PMTs were sent to preamplifiers and then split into two channels: high sensitivity (HS; with additional amplification k ~ 5) and low sensitivity (LS; without amplification). The signals were digitized with 8-channel 12-bit 250-MS/s SIS3320 Struck VME module for the LS channels, and with two 4-channel digital oscilloscopes with GPIB readout: Tektronix DPO7000 (625 MS/s) and LeCroy LT344 (500 MS/s) for the HS channels.

## 2.3. Gas filling and purification system.

The xenon gas for filling the detector was kept in a gas system which consisted of high pressure cylinders, pipes, valves, and getters. All elements of the system were made of stainless steel, and only copper or copper-indium gaskets were used for sealing. Fore-vacuum pumping out of the system and the chamber was performed by oil-free adsorption charcoal and zeolite pumps cooled down by liquid nitrogen. Final pumping out was performed by magnetic-discharge titanium pumps. The process was monitored by a residual gas analyser RGA200. For preliminary purification, homemade hot getters (Ca (700˚C), Ti (900˚C)) were used. Final purification was performed by Mykrolis Megaline purifier during condensation of xenon in the chamber.

## 2.4. Purity issue.

Our major concern during preparation of the detector was possible contamination of the LXe medium by electronegative impurities emanated from organic materials used in the detector. For this reason, we excluded conventional G10 as a material for THGEM and instead chose Kapton. The next most dangerous substance which can contaminate LXe is p-terphenyl. This was prevented by a poly-para-xylilen film. However, the quality of this protection layer has not been studied specifically, and therefore, the probability of having small pin-holes were not excluded. In addition, there was a sufficient amount of Kapton (THGEM and isolating pipes) which is known to not be an ideal material from the point of view of the contamination of the liquid xenon by electronegative impurities. Indeed, we haven't managed to achieve the usual level of outgassing during the pumping out of the detector. The residual pressure dropped dramatically (by about two orders of magnitude) when the pumping was stopped overnight. On contrary, according to our previous experience, the chamber without these new materials usually kept the vacuum at a level of ~ $10^{-5}$ torr for several days.

## 3. Detector operation

During the first filling of the chamber by LXe, we obtained the life time of electrons $\tau_e = 3.3$ +/- 0.9 μs at the operation field equal to 3.75 kV/cm. The life time was measured by averaging a large number of the electroluminescent signals from muon events and fitting them by exponential decay function. Such low value of the electron lifetime was not surprising taking into account potential contamination of LXe by detector components as described above. At subsequent fillings of the detector, we have reached $\tau_e = 10.1 \pm 1.3$ μs.

Typical waveforms recorded from the PMTs and MGPDs are given in figure 3. The event shown in this figure corresponds to a gamma-ray with an energy deposition of ~ 15 keV (480 electrons extracted from the liquid to the gas phase with ~ 7 collected photoelectrons per electron). The red line (I) represents the sum of the waveforms recorded in the LS channel from all PMTs. The blue line (II) represents the sum of the waveforms recorded from all MGPDs. Signal (1) corresponds to scintillation in the LXe; the signal (2) corresponds to electroluminescence in the gas between the LXe surface and the THGEM1 plane; the signal (3), to electroluminescence in the holes of the THGEM1, and the signal (4), to electroluminescence in the holes of the THGEM2. The signal (4) is significantly suppressed due to the limited optical transparency of the THGEM1. The signal (5) on the cumulative waveform (II) is due to the light detected by the MGPDs. Evidently, the appearance time of this signal (5) coincides with that of the signal (4).

The dependence of the MGPD signal on the HV applied to the THGEM2 is shown in figure 4 (waveform 5 in figure 3, triangles in fig. 4) together with the signal from the THGEM1

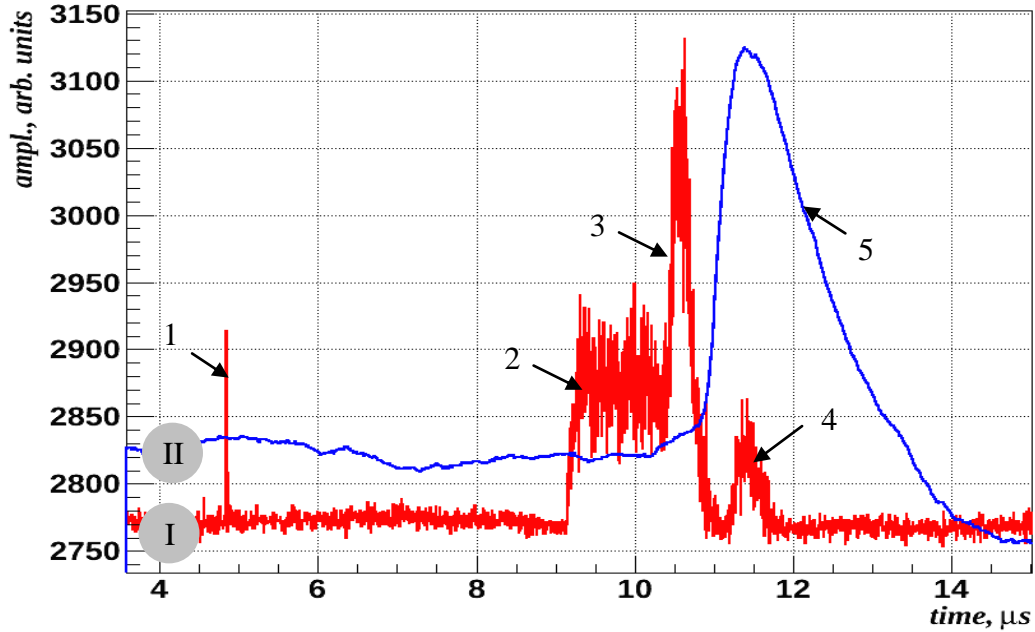

**Figure 3.** Typical sum waveforms from the PMTs (curve I), and from MGPDs (curve II). 1 – scintillation signal in the LXe, 2 – electroluminescence in the gas gap between the LXe surface and THGEM1, 3 – electroluminescence in the THGEM1 ($U_{THGEM1}$=1.6 kV), 4 and 5 – electroluminescence in the THGEM2 ($U_{THGEM2}$=2.1 kV), (*see explanation in text*).

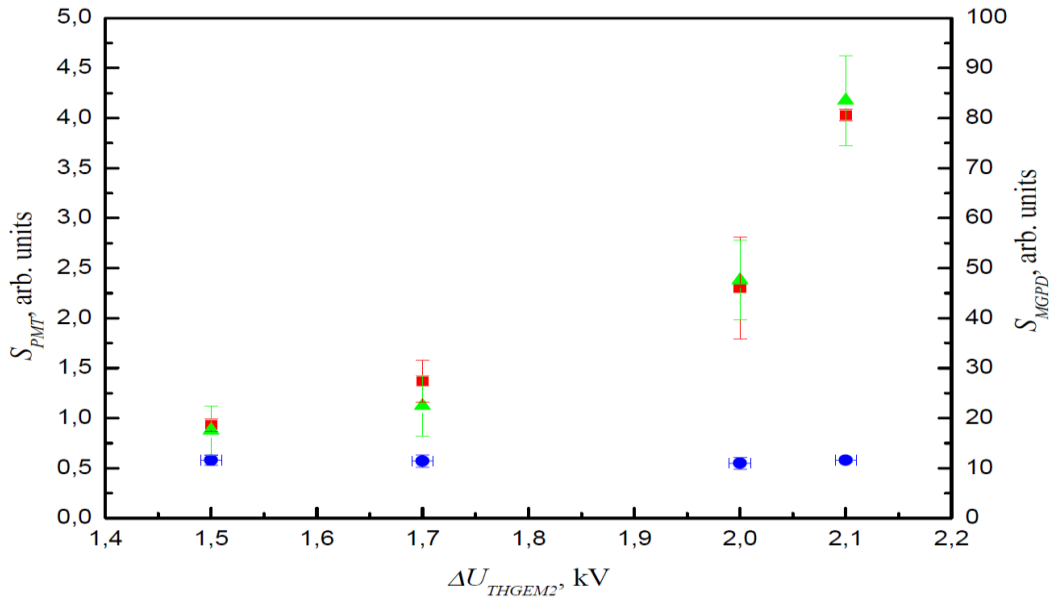

**Figure 4.** The signals from THGEM1 and THGEM2 (areas of the correspondent waveform parts) vs $\Delta U_{THGEM2}$. Circles and squares – recorded by PMTs from THGEM1

recorded by the PMTs (part 3 of the waveform in figure 3, circles in figure 4) and with the signal from the THGEM2 recorded by the PMTs (part 4 in figure 3, squares in figure 4). As we didn't have any low-energy radioactive sources in the LXe volume, and consequently, no monochromatic energy lines, all signals are normalized to the electroluminescence in the gas above the liquid (part 2 in figure 3). The constant behaviour of the signal from the THGEM1 indicates that there was no significant electric field distortion in the THGEM1 caused by the THGEM2. As shown in figure 4 the THGEM2 signal taken from the MGPD array correlates with that of the PMTs, as it should be in the case of the proper performances of both readout systems. It is evident from this figure that we used THGEM at the very beginning of avalanche multiplication. Unfortunately, this type of THGEM did not allow us to test the performance of the detection system at higher voltages due to breakdown problem started at 2.2 kV. On the other hand, our initial concept was to have only electroluminescent amplification in THGEM holes to keep single electron spectroscopy.

The estimated sensitivity of the THGEM+WLS+MGPD readout system to ionisation is $0.75 \pm 0.1$ cells/e.

## 4. Position reconstruction

The important characteristic of the readout system is the accuracy of XY reconstruction. Since we did not have any point-like radioactive source in the LXe volume the only way to estimate the resolution was the comparison of coordinate reconstruction using PMTs and MGPDs as well as the comparison of the experimental data with MC simulations.

In the MC model, the process of photon generation was divided into two parts. In the first one, the VUV photons were generated either in THGEM1 or THGEM2 and transported to the lower hemisphere to the PMTs, taking into account reflection and refraction at the LXe surface. In the second part, the photons were generated in THGEM2 only and traced in the opposite direction. The photons, whose traces ended up at the WLS surface, produced secondary photons with a uniform angular distribution in the upper hemisphere. These photons were detected by the MGPD array.

A centroid XY reconstruction algorithm was applied to the MC generated events. This simple reconstruction algorithm gives the distorted positions of the events vertexes. Since the coordinates of the MC generated events were known in advance, special correction functions between the observed and the real coordinates were obtained for both the PMT and the MGPD arrays. After application of the corrections [27] the distributions of deviations between the known and reconstructed coordinates were obtained for the MC generated events that had energies close to the energies of the selected experimental events (~ 15 keV). These distributions (for Y-coordinate) are shown in figure 5 for the PMT and MGPD arrays *on the left* and *on the right*, respectively. Their spatial resolution (sigma) is equal to $1.026 \pm 0.001$ mm and $0.926 \pm 0.002$ mm, respectively.

Because we did not have a point calibration source in the detector, we cannot compare MC simulated values with the experimental ones separately for the PMT and MGPD arrays. However, we can compare both MC and real data vertexes reconstructed for PMT and MGPD arrays. The difference between Y coordinate reconstructed by PMT and MGPD is shown in

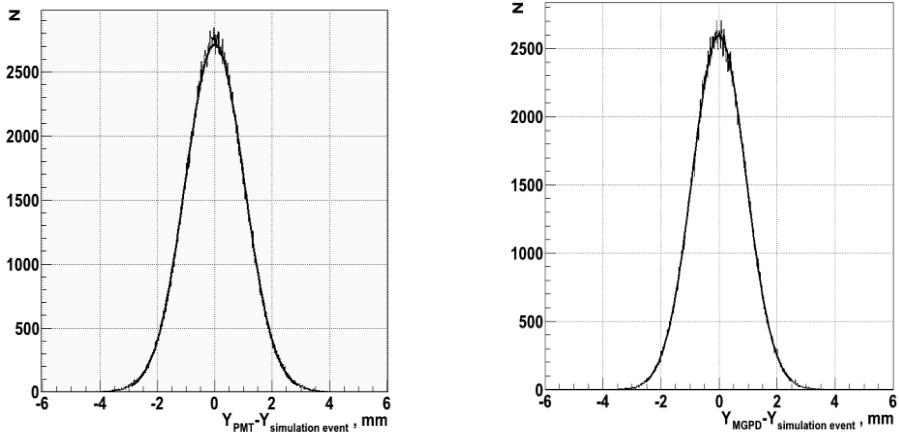

**Figure 5.** Distribution of the deviations between Monte Carlo simulated and reconstructed Y-coordinates for the PMT array (*on left*) and the MGPD array (*on right*).

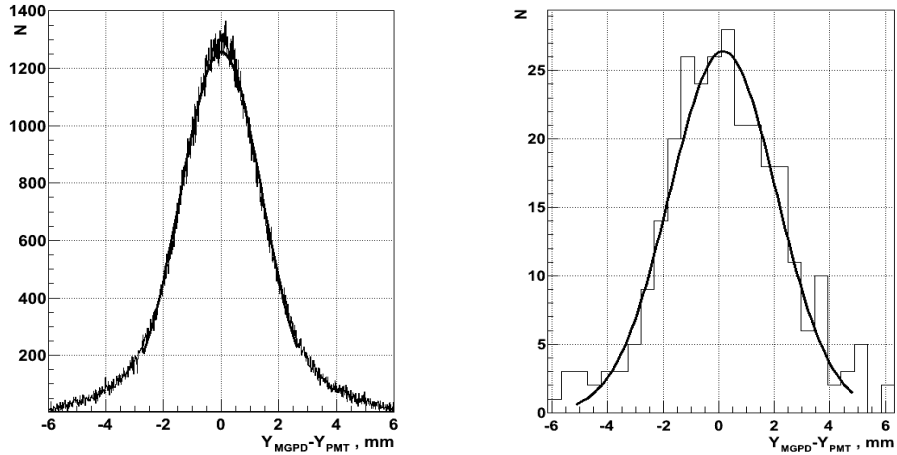

**Figure 6.** Monte Carlo simulated (*on left*) and experimental (*on right*) distribution of the deviations between coordinates obtained by the PMT and MGPD arrays.

figure 6. The sigma for the MC events is 1.357 ± 0.003 mm (*on the left*) and for the experimental data is 1.869 ± 0.147 mm (*on the right*). The experimental distribution is wider than one predicted by simulation but the difference is not significant. We can conclude from this that the spatial resolutions of the PMT and the MGPD arrays are close to each other at a level of 1 mm (sigma). Note, that the number of hit cells per electron for the MGPD array was by an order of magnitude lower than the number of photoelectrons per electron detected in the PMT array. MC simulation demonstrated that the spatial resolution of the MGPD array is significantly better for larger light collection efficiency.

## 5. Discussion

Note that the reached sensitivity of THGEM+WLS+MGPD of 0.75 ± 0.1 cells/e was measured at ~ 1 V overvoltage above the breakdown points of MGPDs because several photodetectors in the array turned out to be very noisy. At this overvoltage, the PDE is only ~ 1/3 of the full value, which is ~ 10% according to our measurements for the combination of the WLS and MGPD [12]. The second issue is the fill factor of the array. For our geometry it is estimated to be ~ 6%. Consequently, with the quite realistic fill factor of ~50%, which can be achieved when the future large area devices will be available and with the full PDE one will have ~ 20 cells/e which is comparable or even higher than we currently have with the PMT array.

At the time when we started designing the optical readout system described in this paper, the only devices available worldwide, which satisfied our requirements, were those by CPTA LTD (blue sensitive ones). Very recently KETEK GmbH [28] has opened a new product line of SiPMs with sensitivity in the blue region. The PDE of the new KETEK SiPMs (> 40% at 420 nm) are higher by a factor of > 2 than the PDE of CPTA devices. Another very recent development is MPPC by Hamamatsu for operation in the LXe environment (for MEG experiment [15]) with a PDE at 175 nm of ~ 10% (~ 15% was reported in [29]).

For the new PM6660 SiPM by KETEK the ratio of the active area (6.0 x 6.0 mm) to the package area (7.0 x 7.5 mm) is 0.69. Thus, with the combination of this photodetector with THGEM+WLS the sensitivity to ionisation of ~ 50 cells/e, can be achieved.

MC simulations predict that the spatial resolution of MGPD array will be significantly better for larger light collection efficiency and probably will be limited by systematic effects.

## 6. Conclusion.

For the first time, the operation of the two-phase xenon emission detector with gas electron photomultiplier and optical readout has been demonstrated using THGEM in combination with MGPD array and wavelength shifter. We measured the yield of 0.75 ± 0.1 cells/e for the current geometry of the THGEM+WLS+MGPD system. Estimation shows that the yield of ~ 50 cells/e can be achieved with the newly developed large area (6.0 x 6.0 mm) devices that have a large active-to-packaging area ratio. The higher yield can be obtained with the higher THGEM gain which is possible with the use of thicker THGEM.

The spatial resolution (sigma) of the MGPD array is ~ 1 mm for ~ 15-keV gamma rays and may be significantly improved for the larger values of light collection efficiency.

The tests have shown the compatibility with the liquid xenon environment of the chosen protection scheme of a p-terphenyl wavelength shifter by a poly-para-xylilen film. Although at the first filling of the chamber not a very high electron lifetime ($\tau_e = 3.3 \pm 0.9$ μs at 3.75 kV/cm) in the liquid xenon was achieved, the following fillings have shown the acceptable purity of the detector media ($\tau_e = 10.1 \pm 1.3$ μs).

**Acknowledgements**

This work was supported by the RFBR grant 09-02-12217-ofi_m, and in part by the Russian Ministry of Education and Science via grants 8174, 8411, 1366.2012.2 and by the RF Government under contracts of NRNU MEPhI with the Ministry of Education and Science of №11.G34.31.0049 from October 19, 2011.

We are very grateful to N.M. Surin and M.Yu. Yablokov, for the development of the wavelength shifter, and to S.A. Zav'yalov for the high quality of manufacturing of the poly-para-xylilen protection film for the wavelength shifter, and to V.N. Solovov for the fruitful discussions.